\pdfoutput=1
 \documentclass[fleqn,usenatbib,usedcolumn]{mnras}
\usepackage[british]{babel}             
\let\la=\lesssim 

\usepackage[T1]{fontenc}
\usepackage{ae,aecompl}
\usepackage{soul}
\usepackage{physics}
\usepackage{steinmetz}

\usepackage{graphicx}	
\usepackage{amsmath}	
\usepackage{amssymb}	

\usepackage{subfig}
\usepackage{float}

\usepackage{color}

 \hypersetup{pdfauthor={E. de Lera Acedo},
               pdftitle={Spectral performance of SKALA antennas I: Mitigating spectral artefacts in SKA1-LOW},
               pdfkeywords={Cosmology, SKA1-LOW, Calibration},
               bookmarksnumbered=true}
\volume{{\rm in press}}

\newcommand{\highlighttext}[1] {#1}

\newcommand{\hltr}[1] {#1}

\title[Spectral performance of SKALA antennas I: Mitigating spectral artefacts in SKA1-LOW]{Spectral performance of \highlighttext{SKA Log-periodic} Antennas I: Mitigating spectral artefacts in SKA1-LOW 21-cm cosmology experiments} 

\author[Eloy de Lera Acedo et al.]{Eloy de Lera Acedo$^{1}$\thanks{E-mail: eloy@mrao.cam.ac.uk},
Cathryn M. Trott$^{2,3}$,
Randall B. Wayth$^{2,3}$,
Nicolas Fagnoni$^{1}$,
\newauthor
Gianni Bernardi$^{4,5}$,
Brett Wakley$^{6}$,
L\'eon V.E. Koopmans$^{7}$,
Andrew J. Faulkner$^{1}$,
\newauthor 
Jan Geralt bij de Vaate$^{8}$
\\
$^{1}$Cavendish Laboratory, University of Cambridge, Cambridge, CB3 0HE, United Kingdom\\
$^{2}$International Centre for Radio Astronomy Research (ICRAR), Curtin University, Bentley Australia\\
$^{3}$Australian Research Council Centre of Excellence for All-Sky Astrophysics (CAASTRO), Australia\\
$^{4}$SKA SA, 3rd Floor, The Park, Park Road, Pinelands, 7405, South Africa\\
$^{5}$Department of Physics and Electronics, Rhodes University, P.O. Box 94, Grahamstown, 6140, South Africa\\
$^{6}$Cambridge Consultants Ltd, Cambridge, United Kingdom\\ 
$^{7}$Kapteyn Astronomical Institute, University of Groningen, P.O.Box 800, 9700AV, Groningen, The Netherlands\\
$^{8}$ASTRON, the Netherlands Institute for Radio Astronomy, Dwingeloo, The Netherlands
}

\date{Accepted XXX. Received YYY; in original form ZZZ}

\pubyear{2016}

\begin{document}
\label{firstpage}
\pagerange{\pageref{firstpage}--\pageref{lastpage}}
\maketitle

\begin{abstract}
This paper is the first \highlighttext{in} a series of papers describing the impact of antenna instrumental artefacts on the 21-cm cosmology experiments to be carried out by the \highlighttext{low frequency instrument (SKA1-LOW) of the Square Kilometre Array telescope (SKA)}, i.e., the Cosmic Dawn (CD) and the Epoch of Reionization (EoR). The smoothness of the passband response of the current log-periodic antenna being developed for the SKA1-LOW is analyzed using numerical electromagnetic simulations. The \highlighttext{amplitude variations over the frequency range} are characterized using low-order polynomials defined locally, in order to study the impact of the passband smoothness in the instrument calibration and CD/EoR Science. A solution is offered to correct a fast ripple found at 60~MHz during a test campaign at the SKA site at the Murchison Radio-astronomy Observatory, Western Australia in September 2015 with a minor impact on the telescope's performance and design. A comparison with the Hydrogen Epoch of Reionization Array antenna is also shown demonstrating the potential use of the SKA1-LOW antenna for the Delay Spectrum technique to detect the EoR. 
\end{abstract}




\section{Introduction}

\begin{figure*}
\centering
\includegraphics[width=3.4in]{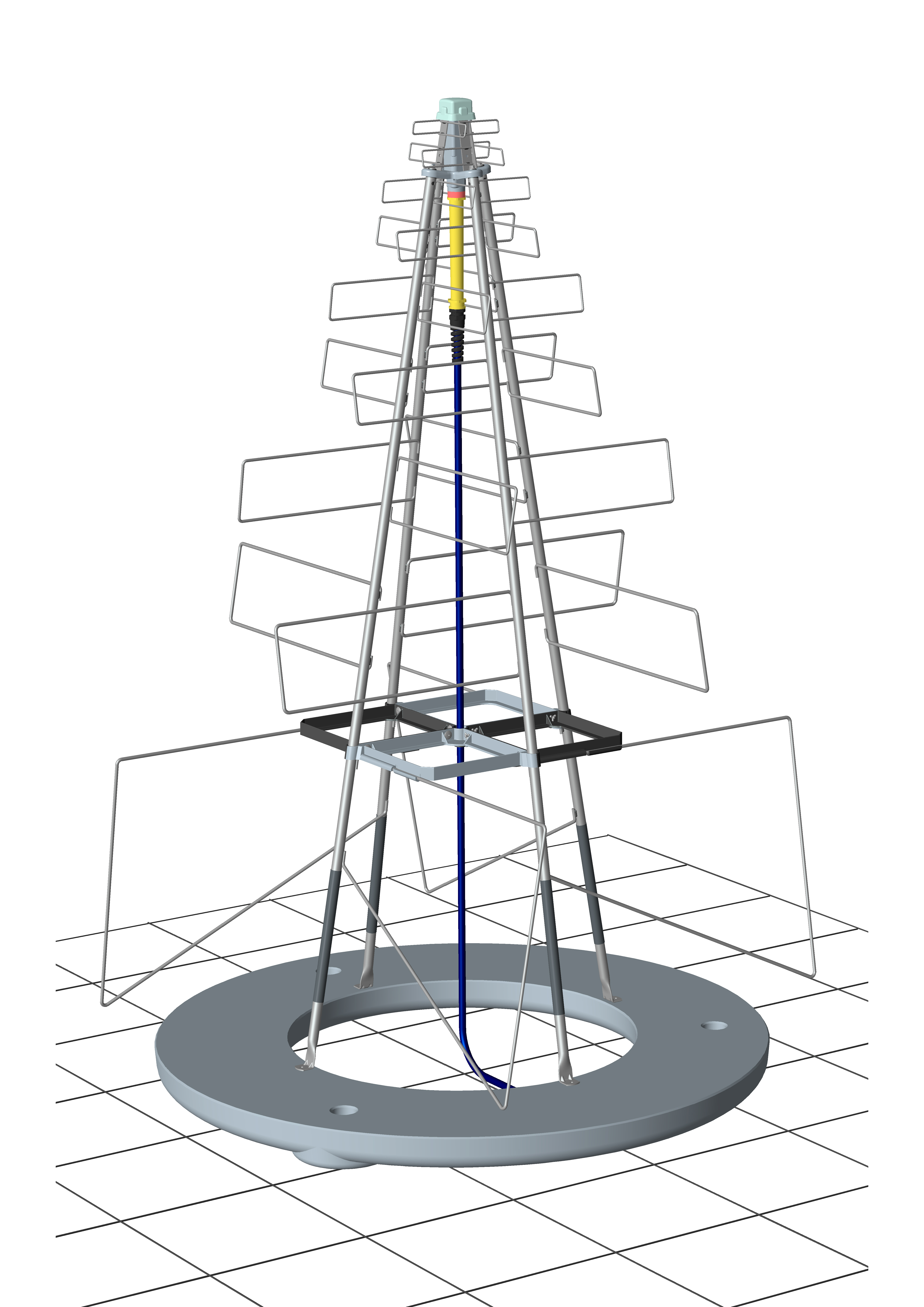}
\caption{Computer model of SKALA-2 antenna on top of a metallic ground plane (the ground plane is a mesh of wires with 30 cm pitch). The antenna is a log-periodic dipole array made of 4 identical metallic arms forming 2 polarizations with 9 dipoles each. The gray ring is the base of the antenna. The cabling coming down vertically in the center of the antenna \highlighttext{is} the optical fiber and power cable. The top enclosure with a green cap hosts the low-noise amplifiers and \highlighttext{an} RF-over-Fiber transmitter. The antenna has a footprint of 1.2 x 1.2 m and a height of 1.8 m.}
\label{antenna}
\end{figure*} 

The Square Kilometre Array (SKA)\footnote{http://www.skatelescope.org} is \highlighttext{a} next generation radio telescope, with unprecedented sensitivity and survey speed. The SKA will explore the southern hemisphere skies faster and deeper than any other radio telescope has ever done. Spread across two continents (Southern Africa and Australia + New \highlighttext{Zealand}), it will cover a frequency range from 50 MHz up to 14 GHz and will tackle fundamental problems in modern Astrophysics including the study of the Cosmic Dawn (CD) and the Epoch-of-Reionization (EoR) as described in~\citet{Koopmans2015}. Phase I of the SKA will have a lower frequency instrument (SKA1-LOW: 50-350 MHz) that will be used for the study of the early epochs of the Universe by looking at the redshifted signal from the 21-cm Hydrogen emission line. This instrument will be located in Western Australia. It will be a 512 station interferometer with baselines up to 60 km and more than 130,000 antenna elements \citep{Turner2015}. Each station has been \highlighttext{designed} as a pseudo-random array of wide band antennas in order to minimize the effects of mutual coupling as shown by \citet{Gonzalez2011} and \citet{deLera2011b} and the effects of side-lobes \citep{Razavi2012,El-makadema2014}. Each station is 35~m in diameter in order to meet the desired beam-width \citep{Mellema2013,Koopmans2015}. 

The SKA1-LOW antenna is \highlighttext{proposed to be} a log-periodic dipole array, SKALA (SKA Log-periodic Antenna). \highlighttext{Version 1 of SKALA, SKALA-1 was described in \citet{deLeraAcedo2015}. SKALA-1} maximizes the telescope's sensitivity over the field of view across the 7:1 frequency band due to its moderate to high directivity and flat impedance, \highlighttext{in contrast to dipole antennas} that \highlighttext{have low directivity values and only show relatively flat impedance over a narrow frequency band (typically 2:1 or 3:1 at most)}. \highlighttext{Subsequently, a mechanical upgrade of SKALA-1 was designed and built, SKALA-2, in order to improve the design for mass manufacturing with negligible impact on the electromagnetic performance of the antenna. This upgraded design was presented in \citet{deLeraAcedo2015b}. A computer model of the \highlighttext{SKALA-2} antenna is shown in Figure~\ref{antenna}. This paper describes the modifications to the length of the SKALA-2 antenna arms and component values of capacitors and inductors in the input matching network of the Low-Noise Amplifier (LNA) in order to improve the spectral smoothness of the antenna system. The result of this improvement is SKALA-3.}

Both the CD and EoR observations require a smooth response of the instrument across frequency to enable the detection of the faint signals buried in much brighter \highlighttext{foreground signals} and not to introduce instrumental spectral structure on scales of relevance for the neutral hydrogen signal \citep{Trott2016,Parsons2012, DeBoer2016}.

This paper describes the analysis of the \highlighttext{proposed} SKA1-LOW antenna, \highlighttext{SKALA-3}, using a local low-order polynomial fitting in order to assess its performance against realistic CD/EoR science requirements. In Section~\ref{sec:requirements} the requirements derived in \citet{Trott2016} are summarized. Section~\ref{sec:antenna} describes the current antenna topology and performance. Section~\ref{sec:analysis} focuses on the description of the analysis and the results from numerical simulations. Section~\ref{sec:optimisation} presents \highlighttext{the improved bandpass smoothness to meet the science requirements} of the \highlighttext{proposed} SKA1-LOW antenna. Section~\ref{sec:HERA} shows a comparison with the Hydrogen Epoch of Reionization Array (HERA\footnote{http://reionization.org}) antenna against requirements derived for the Delay Spectrum technique \highlighttext{for EoR observations}. Finally, Section~\ref{sec:conclusions} draws some conclusions.

\begin{figure*}
\centering
\includegraphics[width=4in]{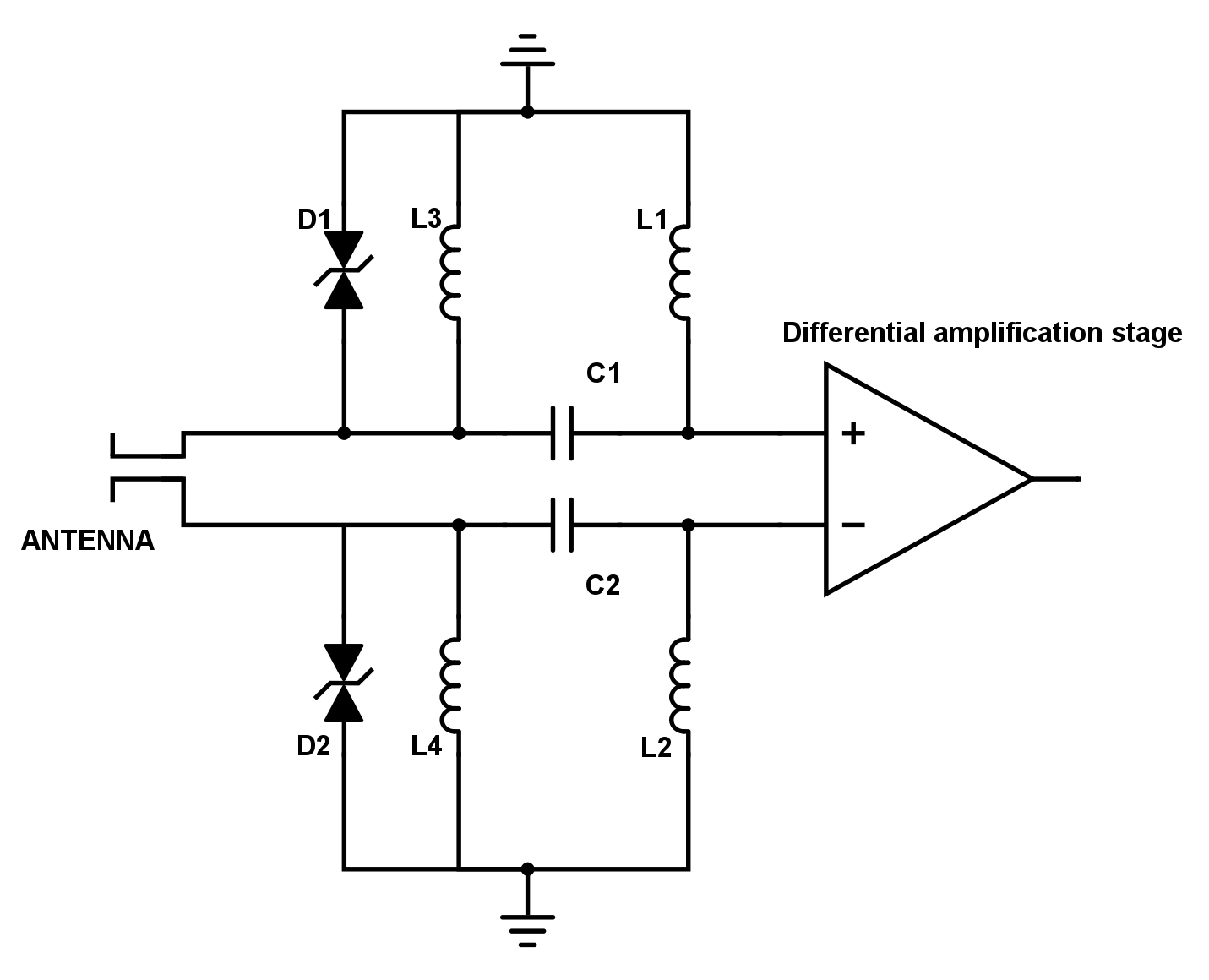}
\caption{\highlighttext{Simplified RF schematic model of the first amplification stage of the LNA. The S-parameters and noise parameters for the transistors used in this design are obtained from the manufacturers. This diagram highlights the key components in the input side of the LNA that have been used for the improvement of the design (the DC blocking capacitors C1 and C2, the bias inductors L1, L2 and the ESD shunt inductors L3 and L4). The ESD diodes D1 and D2 are also shown here. For the simulations in this paper the full model of the LNA also included the second stage transistor although this has a small effect on the matching with the antenna.}}
\label{fig:RF_simulation}
\end{figure*}

\begin{figure*}
\begin{minipage}{.5\linewidth}
\centering
\subfloat[]{\label{main:a}\includegraphics[width=3in]{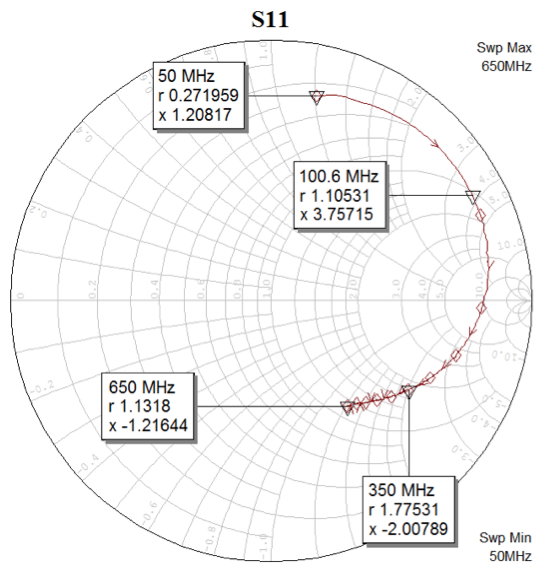}}
\end{minipage}%
\begin{minipage}{.5\linewidth}
\centering
\subfloat[]{\label{main:b}\includegraphics[width=3in]{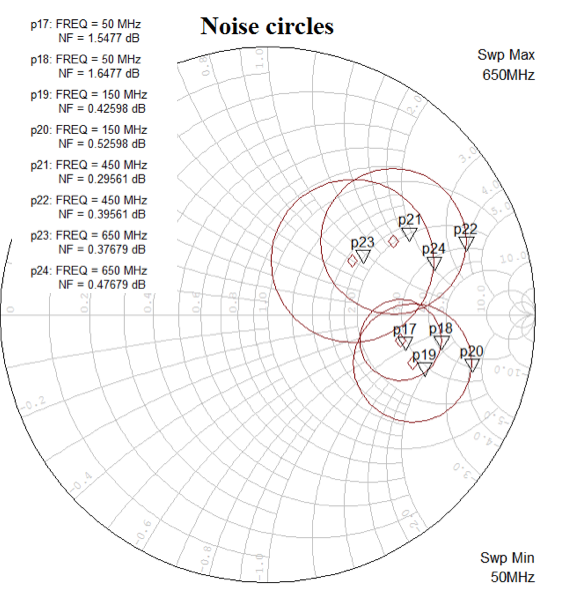}}
\end{minipage}\par\medskip
\centering
\subfloat[]{\label{main:c}\includegraphics[width=4in]{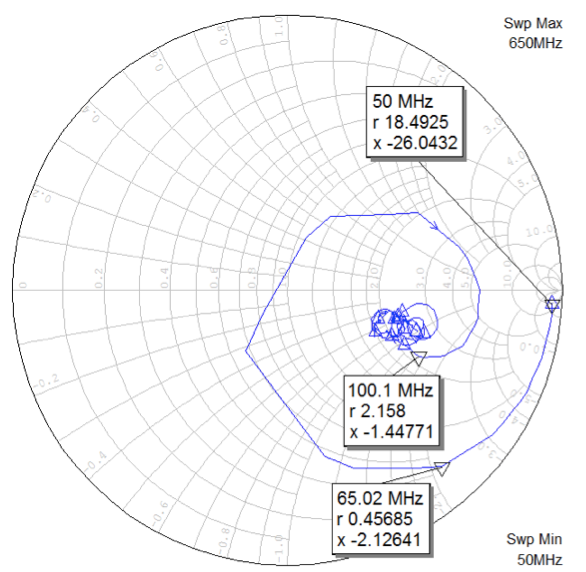}}
\caption{Antenna and LNA impedances; (a) LNA input impedance for power match, (b) LNA noise circles for noise match and (c) Antenna input impedance. \highlighttext{The Smith chart is plotted on the complex reflection coefficient plane in two dimensions. The horizontal axis corresponds to a purely real reflection coefficient and the center is the reference impedance, 50 $\Omega$.}}
\label{fig:antennaLNAmatch}
\end{figure*}

\section{Spectral calibration requirements}
\label{sec:requirements}

In \citet{Trott2016} a formal procedure for the derivation of spectral requirements for radio interferometers dedicated to 21-cm cosmology experiments is presented. Here, we use the requirements derived in that paper to assess the design of the \highlighttext{SKALA-2 and SKALA-3 antennas}. In a companion paper \citep[Paper II]{trottskala17}, we use realistic simulations to test the calibration performance of \highlighttext{these antennas}, and a comparison dipole \citep[the Murchison Widefield Array \highlighttext{- Engineering Development Array}]{tingay13_mwasystem,wayth16}. 

\begin{table}
	\centering
	\caption{Requirements from~\citet{Trott2016}. $\delta$ is the normalized fractional residual after \highlighttext{a third-order} polynomial fitting.}
	\label{tab:requirements}
	\begin{tabular}{lccr} 
		\hline
		Frequency (MHz) & $\delta$ \\
		\hline
		50 & 0.025\\
		100 & 0.01\\
		150 & 0.005\\
		200 & 0.008\\
		\hline
	\end{tabular}
\end{table}

One of the \highlighttext{spectral} requirements affects the EoR Power spectrum pipeline and is related to the level of the amplitude residuals after fitting a low-order polynomial locally to the station's complex-valued voltage gain passband. \highlighttext{The power derived from fitting a low-order ({\it n} = 2, 3, 4) polynomial} over three contiguous coarse channels (2.25 MHz) yields structured power in the power spectrum space. This power is less than the thermal noise, but has the potential to cause low-level bias in the derived science results due to its shape. Table~\ref{tab:requirements} lists the relative fractional bandpass \highlighttext{errors} tolerable in a fourth-order polynomial residual on voltage gain amplitudes at 4 different frequencies \citep[In][requirements were also derived for second and fourth other polynomial fittings]{Trott2016}. \highlighttext{Strictly speaking this should be applied to the station beam passband. Our} current analysis has been done at the antenna level, which will dominate the passband response of the instrument. Further analysis is needed to include station beam effects such as mutual coupling and other antenna \highlighttext{effects} such as polarization and the instrument's response across the whole field of view (including spectral variations in the far field pattern of the antenna). 

Furthermore, in \citet{Trott2016} requirements are derived for the necessary smoothness in the phase passband of the station's voltage gain. Phase residuals affect the imaging (tomography) experiment substantially. The so-called ``phase noise" decorrelates the visibilities and adds an effective noise to the images that swamps the signal. So, in \citet{Trott2016} the maximum change in phase across any fine channel (within the 100 - 200 MHz band) that would lead to swamping of the expected EoR signal is computed. In the worst case where the phase residual on each channel is fixed over the full experiment, there \highlighttext{cannot} be more than a 0.04 degree change of phase over any fine channel (4.58 kHz). An optimistic case where the phase residual is uncorrelated in time between calibration cycles, yielding an increase in dynamic range and weaker constraints, \highlighttext{raises} that limit to 0.2 degrees. 

\section{SKALA antenna and LNA performance}
\label{sec:antenna}

The \highlighttext{SKALA-1 and SKALA-2} antennas were designed to maximize the sensitivity of the SKA1-LOW instrument across at least a 7:1 frequency band by means of optimizing its effective area, minimizing \highlighttext{its} footprint and average distance between elements (to maximize brightness sensitivity for a fixed number of elements) and reducing receiver noise (dominated by the matching between the antenna and the first stage low-noise amplifier). 

\begin{figure}
\includegraphics[width=\columnwidth]{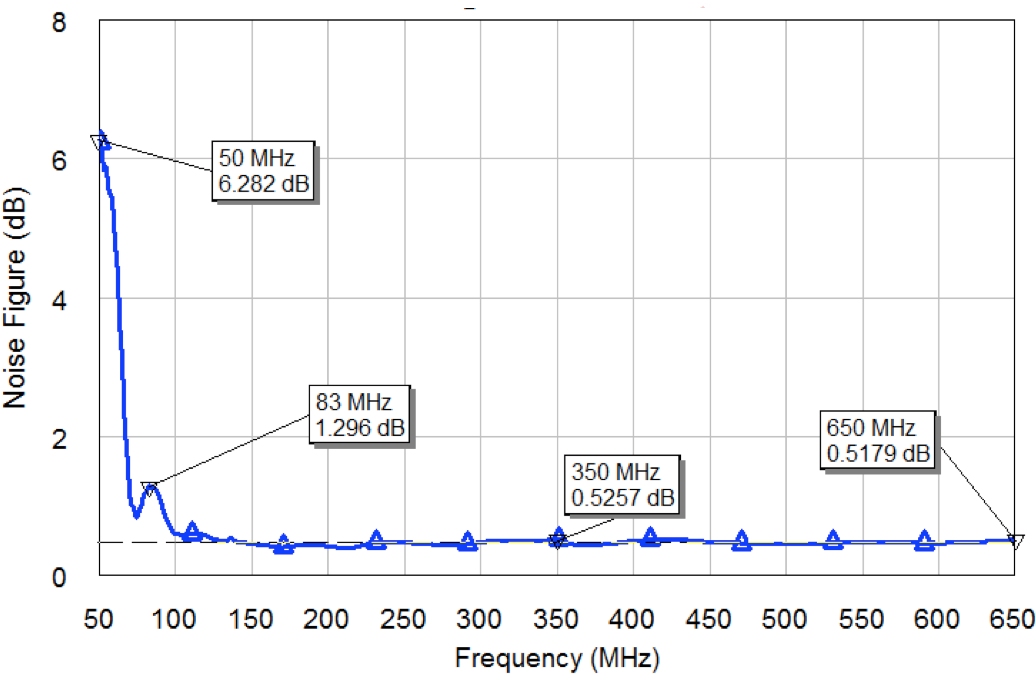}
\caption{Noise figure of LNA when connected to SKALA-2.} 
\label{fig:NFori}
\end{figure}

The antenna simulations were carried out using CST\footnote{Computer Simulation Technology AG, http://www.cst.com} for the antenna and Microwave Office\footnote{http://www.awrcorp.com/products/ni-awr-design-environment/microwave-office} for the LNA. \highlighttext{The LNA design is based on a pseudo-differential configuration with a first differential amplification stage using Qorvo\footnote{http://www.qorvo.com} transistors (TQP3M9039) and a second amplification stage after a wide band balun transformer using a Mini-Circuits\footnote{http://www.minicircuits.com} transistor (PSA-5451+).} Figure~\ref{fig:RF_simulation} shows the \highlighttext{simplified} schematic of the first amplification stage. Figure~\ref{fig:antennaLNAmatch} shows the input LNA impedance ($Z_{LNA}$), the LNA optimum noise impedance at the center of the noise circles ($Z_{opt}$) and the antenna input impedance ($Z_{A}$) across frequency. It is well known that the process of noise matching and power matching \highlighttext{calls} for a trade-off between achieving a low noise figure and a smooth passband with maximum power transfer~\citep{Pozar2011microwave}. Figure~\ref{fig:antennaLNAmatch} shows \highlighttext{that} the optimum noise impedance of the LNA, which we need to match our antenna to for minimum noise figure, and the input impedance of the LNA, which we need to match our antenna impedance to for maximum power transfer, are not in the same place in the Smith chart. \highlighttext{SKALA-1} was originally designed to cover the band of 70-450 MHz and it was later optimized to work down to 50 MHz with the focus on maximum sensitivity and therefore minimum receiver noise, hence \highlighttext{it is matched} to the optimum noise impedance of the LNA. \highlighttext{The noise figure of the LNA connected to SKALA-2 is} shown in Figure~\ref{fig:NFori} and the passband \highlighttext{is} shown in red in Figure~\ref{fig:passband_amp_comp} and Figure~\ref{fig:passband_phase_comp}. The \emph{passband} here is the voltage gain of the LNA, $G_{LNA}$, when connected to the antenna impedance (\highlighttext{not} a fixed reference impedance), thus \highlighttext{the performance is} as it would be in a real scenario. The antenna was simulated standing on top of an infinite perfect electric conductor ground plane. A low receiver noise was achieved down to 100 MHz where the sky noise starts to severely dominate the system noise and we can relax the receiver noise requirements. However, the passband response of the LNA when connected to the antenna impedance, exhibits several sharp features, especially one around 60 MHz. 

It is important to notice that according to the definition of sensitivity used here and described by equation~\ref{eq:AoT} (effective area over system temperature or signal over noise) the \highlighttext{the passband smoothness will not be visible in that metric}. \highlighttext{The reason for this is that the ratio of effective area over system noise temperature is constant independent of where the ratio is taken in the receiving chain. In equation~\ref{eq:AoT} we refer this ratio to the input of the LNA and therefore both effective aperture and system noise temperature (all of the noise terms) will be equally weighted by the gain of the LNA, as if we try to calculate this ratio at the output of the LNA. This is correct as long as we have taken into account the noise contribution from the rest of the receiving chain (after the LNA) weighted by the gain of the previous stages (in order to refer it to the input of the LNA). Since the gain of the LNA is high and the noise figure of the following stages is small, this contribution is negligible. Consequently, the effect of the passband gain of the LNA will not be visible in the sensitivity values}. In equation~\ref{eq:AoT}, $\lambda$ is the wavelength, $D_{\theta ,\psi }$ is the antenna's directive gain, $\eta$ is the radiation efficiency, $T_{A}$ is the antenna temperature (including the sky temperature), $T_{0}$ is the ambient temperature (295 K) and $T_{rec}$ is the receiver noise temperature. For more details, see \citet{deLeraAcedo2015}.  

\begin{equation}
\left (\frac{A_{eff}}{T_{sys}}  \right )_{\theta ,\psi }=\frac{\frac{\lambda ^{2}}{4\pi }D_{\theta ,\psi }\eta }{\eta T_{A}+(1-\eta )T_{0}+T_{rec}}
\label{eq:AoT}
\end{equation}

\begin{figure*}
\centering
\includegraphics[width=7in]{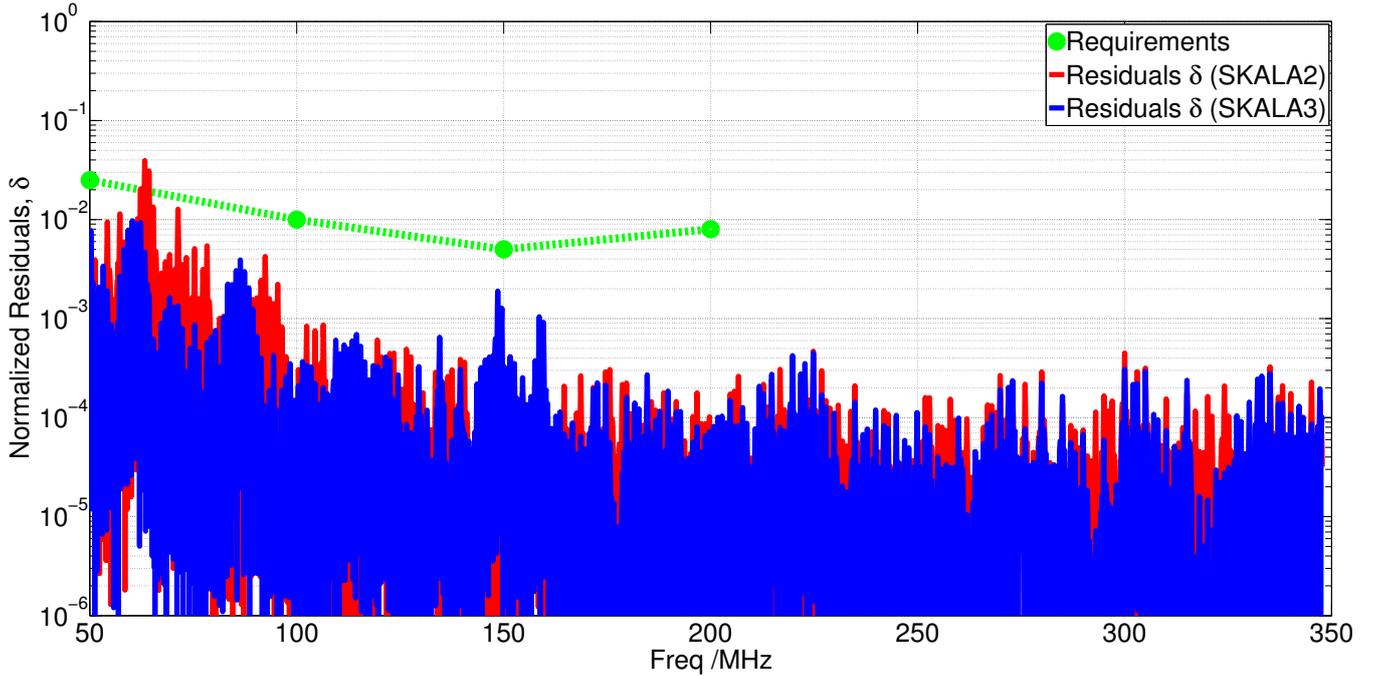}
\caption{Residuals after a 3rd order local polynomial fitting using SKALA-2 \hltr{and SKALA-3}.}
\label{fig:Res_ori}
\end{figure*}

\begin{figure*}
\centering
\includegraphics[width=7in]{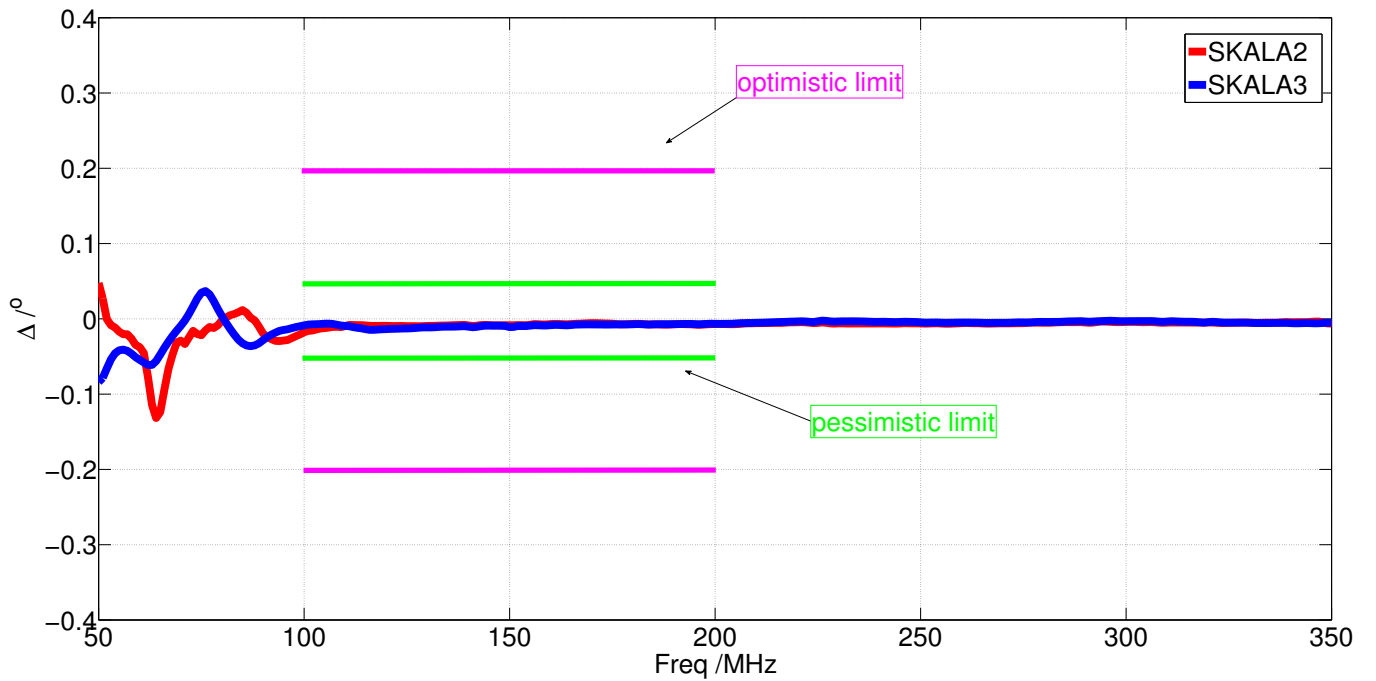}
\caption{Phase gradient per fine channel for SKALA-2 \hltr{and SKALA-3}. The plot also shows the \textit{optimistic} and \textit{pessimistic} limits as defined in \citet{Trott2016}.}
\label{fig:phase_gra_ori}
\end{figure*}

\section{Spectral analysis using local low-order polynomial fitting}
\label{sec:analysis}

In this section we analyze the response of the \highlighttext{SKALA-2} antenna measuring it against the requirements presented in \citet{Trott2016}. As mentioned before we simulate the passband of one antenna alone, because it dominates the passband response across frequency of the station beam. Furthermore, we focus on the passband response of the LNA when connected to the antenna omitting the effect of the antenna far field pattern since this is much more smooth than that of the LNA and will not affect these results. \highlighttext{The frequency band (50 - 350 MHz) is subdivided into coarse channels 750 kHz wide, each of which is subdivided into fine channels 4.6 kHz wide. This means 163 fine channels per every coarse channel}. We fit an {\it n}-order polynomial (by default in this paper the order is 3) to all \highlighttext{fine} frequency channels \highlighttext{in a least-squares sense} across {\it m} coarse channels (by default in this paper {\it m} is 3 as well) to both the real and imaginary parts of the passband. \highlighttext{The fitted \hltr{complex passband ($S^{*}$), where the fit has been performed separately to its real part ($S_{r}^{*}$) and its imaginary part ($S_{i}^{*}$),} follows the definition of equation ~\ref{eq:poly} ($\nu$ is the frequency and $\nu _{0}$ is the central frequency for each coarse channel).} The polynomial fit is only used to model \highlighttext{the fine channels corresponding} to the middle coarse channel of the {\it m} channels used to generate the fit. We can then subtract the model of the amplitude response from the \highlighttext{amplitude of the passband data ($S$) and normalize this difference by $S$ again to get the residuals ($\delta$)} as in equation ~\ref{eq:res}. This process is repeated for all the \highlighttext{fine channels in the frequency bandwidth by moving the central coarse channel by one (and consequently fitting the next set of fine channels)} to obtain the wide-band performance of the antenna. 
\hltr{While the default bandwidth used for the polynomial fitting in this paper is 2.25 MHz (3 coarse channels), in \citet{Barry2016} it is recommended that smoothness is also achieved in larger bandwidths in order to prevent contamination of the relevant modes for the EoR detection. We have therefore extend our analysis here to 9 coarse channels (6.75 MHz). 
The choice of the order of the polynomial is also important, since high orders could constrain the fit in excess and remove not only the instrument's passband but also the cosmological signal. Furthermore, high-order polynomial fittings would result as well in high-order residuals at potentially similar spectral scales to that of the cosmological signal. While he have stayed at a relatively low order here (3), in \citet{trottskala17} the effects of using higher order polynomials are also discussed.}

\begin{equation}
S_{r,i}^{*}(\nu)=A(\nu -\nu _{0}){^{3}} + B(\nu -\nu _{0}){^{2}} + C(\nu -\nu _{0}){^{1}} + D
\label{eq:poly}
\end{equation}

\begin{equation}
\delta(\nu) = \frac{\left |\left |S^{*}(\nu)  \right | - \left |S(\nu)  \right |  \right |}{\left |S(\nu)  \right |} 
\label{eq:res}
\end{equation}

\highlighttext{The phase gradient per fine channel, $\Delta(\nu)$, is calculated using equation ~\ref{eq:phaslope}.}

\begin{equation}
\Delta(\nu) = \dv{ \phase {S(\nu)} }{\nu}*4.6 \text{kHz} 
\label{eq:phaslope}
\end{equation}

Figure~\ref{fig:Res_ori} shows how the required residuals are not met \hltr{by SKALA-2} at the frequency where the passband has peaked behaviour in amplitude and a fast slope in the phase due to the mismatch between the antenna and the LNA. The phase gradient requirement seems to be more under control than the amplitude requirement, especially in the band 100 to 200 MHz, although it is again higher at lower frequencies as indicated in Figure~\ref{fig:phase_gra_ori}.
 
\section{Improvement of the bandpass smoothness}
\label{sec:optimisation}

\subsection{Antenna and LNA modifications}

The spectral response of the antenna is dominated by the mismatch between the antenna and the first stage low-noise amplifier. In order to \highlighttext{improve} the power match between the antenna and the low-noise amplifier, we have explored modifications of both systems \highlighttext{specifically} where we were not meeting the passband requirements (50-100 MHz). 

\highlighttext{In SKALA-3}, we have modified the input network of the LNA (see Figure~\ref{fig:RF_simulation}) using the original input matching components but with a lower value of the DC blocking capacitors and bias inductors \highlighttext{(150 nH ESD shunt inductors remain the same (L3 and L4), 24 pF DC blocking capacitors (C1 and C2), 270 nH bias inductors (L1 and L2))}. The trade-off is \highlighttext{a} higher noise figure, peaking at almost 8.6 dB at 57 MHz. This is a side effect of using a smaller-value inductor in the bias network after we added a resistor for stability. We don't want to change that resistor, and making that inductor larger immediately \highlighttext{sharpens} the peak in the gain response. The series capacitor \highlighttext{could} be reduced in value to smooth the ripple even further, but the peaking in the noise figure starts to increase rapidly (18 pF gives \highlighttext{in} excess of 10 dB noise figure). 

\begin{figure}
\includegraphics[width=\columnwidth]{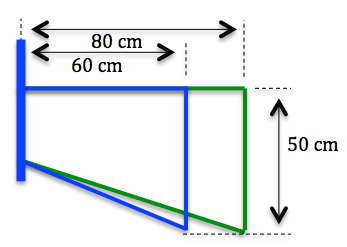}
\caption{New lower antenna arm (green, SKALA-3) versus old arm (blue, SKALA-2).} 
\label{fig:NewArm}
\end{figure}

In order to bring the antenna impedance closer to the LNA input impedance, facilitating the match, we have enlarged the bottom dipole of the antenna (active at the lower end of the band). \highlighttext{The design} is shown in Figure~\ref{fig:NewArm}. This \highlighttext{increases} the footprint of the antenna \highlighttext{to} 1.6 x 1.6 m instead of 1.2 x 1.2 m. While it is possible from an electromagnetic point of view to interleave antenna footprints in a SKA station it may not be \highlighttext{appropriate} for maintenance of the antennas. We are exploring a larger station size (40 m) to support this \highlighttext{new design} while minimizing the impact in the station performance.

\begin{figure}
\includegraphics[width=\columnwidth]{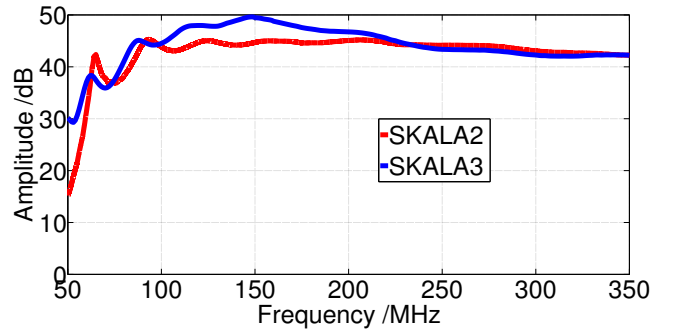}
\caption{\highlighttext{Amplitude of the LNA voltage passband when connected to SKALA-3 (blue) vs when connected to SKALA-2 (red). The plot shows the sharp feature found in the passband of SKALA-2 at 60 MHz.}} 
\label{fig:passband_amp_comp}
\end{figure}

\begin{figure}
\includegraphics[width=\columnwidth]{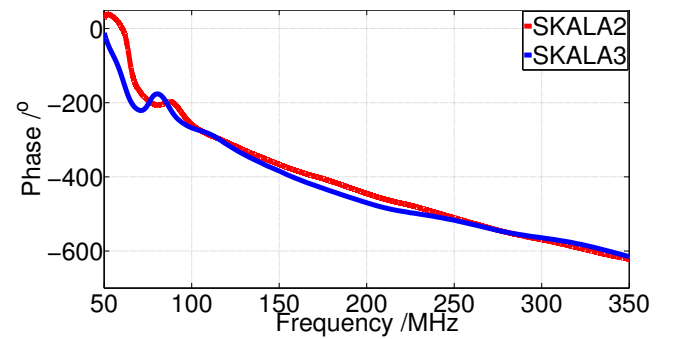}
\caption{\highlighttext{Phase of the LNA voltage passband when connected to SKALA-3 (blue) vs when connected to SKALA-2 (red).}}
\label{fig:passband_phase_comp}
\end{figure}

%

\subsection{Improved antenna response and impact on SKA1-LOW sensitivity}

\begin{figure*}
\includegraphics[width=7in]{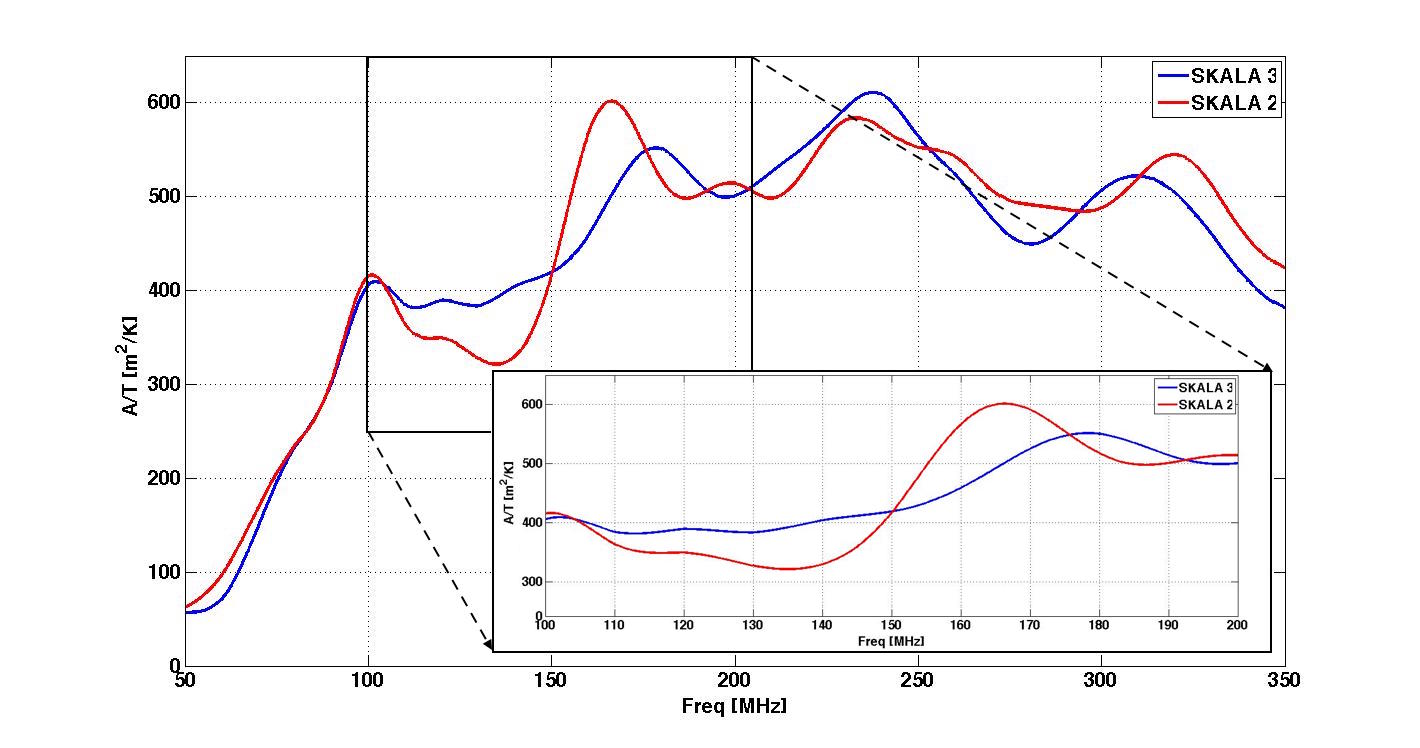}
\caption{SKA1-LOW sensitivity at zenith with SKALA-2 and with SKALA-3 across the full band. Both calculations assume the same parameters (sky noise, station size, etc.) as described in~\citet{deLeraAcedo2015}.} 
\label{fig:AoTfull}
\end{figure*}

With the changes presented above we recalculated the performance \highlighttext{of the new version of the SKALA antenna (SKALA-3)}. Figures ~\ref{fig:passband_amp_comp} and ~\ref{fig:passband_phase_comp} show the new LNA amplitude and phase passbands when connected to the antenna impedance. Figure~\ref{fig:Res_ori} shows the amplitude residuals \hltr{for SKALA-3}, now within the requirements, for a third order polynomial fitted across 3 coarse channels. In Table ~\ref{tab:channels} we also list the residuals for wider bandwidths and a polynomial of order 3 fitting the data. We simulated for 5 coarse channels (3.75 MHz), \hltr{7} coarse channels (5.25 MHz) and 9 coarse channels (6.75 MHz). We observe an increase in the residuals when we try to fit a wider frequency bandwidth with a polynomial of order 3 but always staying below the \highlighttext{requirement limit}.  Figure~\ref{fig:phase_gra_ori} shows how the phase gradient has improved at all frequencies therefore still meeting the requirements in the EoR band. Finally, Figure~\ref{fig:AoTfull} show the effect on the SKA1-LOW sensitivity when using the new antenna, SKALA-3. The sensitivity has been calculated using equation~\ref{eq:AoT}. This is also flatter across frequency. The antenna's directive gain (directivity) is one of the main contributors to the effective aperture and therefore to the sensitivity of the antenna as described in \citet{deLeraAcedo2015}.  The change in the length of the bottom dipole of SKALA in order to improve the low frequency impedance match with the LNA has caused, as expected, a smoother transition between the bottom dipole and the second dipole. This has produced a flatter directivity of the antenna across frequency and therefore a flatter sensitivity pattern. 

\begin{table}
	\centering
	\caption{\highlighttext{Residuals obtained when fitting wider bandwidths.}}
	\label{tab:channels}
	\begin{tabular}{lccr} 
		\hline
		Frequency (MHz) & $\delta$(m = 5) & $\delta$(m = 7) & $\delta$(m = 9) \\
		\hline
		50 & 0.00073 & 0.00612 & 0.01327\\
		100 & 0.00009 & 0.00017 & 0.00074\\
		150 &  0.00090 & 0.00100  & 0.00110\\
		200 &  0.00003 & 0.00003 & 0.00004\\
		\hline
	\end{tabular}
\end{table}

\section{The Delay Spectrum metric}
\label{sec:HERA}

A well-studied way to look at antenna spectral requirements is from the perspective of foreground {\it avoidance} in power spectrum space. In the avoidance scheme, smooth--spectrum foregrounds should, in the ideal case, occupy a wedge--shaped region of the two--dimensional power spectrum space \highlighttext{(where the wave number $k$ can be decomposed into $k_\perp$ and $k_\|$ that are the transverse and line-of-sight wave numbers respectively)} whereas the remaining area -- the so--called EoR window -- is dominated by the 21~cm emission \citep{Datta2010,Morales2012,Thyagarajan2013,Liu2014a,Liu2014b,trott12,Vedantham2012}. \highlighttext{Sources have most of their emission at low $k_\parallel$ values although, due to the inherent chromatic interferometric response, this area increases with baseline length \highlighttext{to resemble} a characteristic wedge--like shape \citep{Parsons2012,Liu2014a,Liu2014b,Thyagarajan2015,trott12,Vedantham2012}.} In the most optimistic scenario, the EoR power spectrum can be directly measured in the EoR window, whose boundaries are set by the so called {\it horizon limit}, i.e. the maximum delay that an astrophysical signal can experience, given by the separation of the two receiving elements. In practice, the boundaries of the EoR window can be narrowed by a number of mechanisms that spread power from the foreground dominated region into the EoR window, in particular calibration errors \citep{Barry2016,Patil2016}, leakage of foreground polarization \citep{Bernardi2010,Jelic2010,Moore2013,Asad2015,Asad2016} and intrinsic chromaticity of the instrumental response. Recent attention has indeed \highlighttext{been} given to simulate and characterize the element response, particularly for the Hydrogen Epoch of Reionization Array \citep{DeBoer2016}, realizing that it may be one of the critical items responsible \highlighttext{for spilling} power from the wedge into the EoR window.

The HERA design leverages a highly-redundant array configuration that accumulates its maximum sensitivity on a limited number of $k_\perp$ modes, building on the results of the Precision Array to Probe the Epoch of Reionization \citep[PAPER]{Parsons2014,Jacobs2015,Ali2015}. Like PAPER, HERA plans to employ a foreground avoidance scheme on a per-baseline base, through the delay transform. In equation ~\ref{eq:delay_1}, $\widetilde{V_{b}}(\tau)$ is the delay ($\tau$) transform (inverse Fourier transform) of the measured frequency visibility for a fixed baseline $V_{b}(\nu )$ (equation~\ref{eq:delay_2}), that includes the sky brightness, $I(\widehat{s},\nu )$ and the antenna's directional power pattern, $A(\widehat{s},\nu )$. $c$ is the speed of light, $b$ is the baseline vector and $\widehat{s}$ is the direction on the sky (unit vector). The sky power spectrum  $P(k)$ can then be approximated by equation~\ref{eq:delay_3}, where $B$ is the effective bandwidth, $\Omega_{b}$ is the integrated beam response, $k_{B}$ is the Boltzmann's constant and $X$ and $Y$ are cosmological parameters relating angular size and spectral frequency to cosmic volumes respectively. For more details, see \citet{Parsons2012}.

\begin{equation}\label{eq:delay_1}
\widetilde{V_{b}}(\tau )\equiv \int V_{b}(\nu)e^{i2\pi\nu\tau }d\nu
\end{equation}

\begin{equation}\label{eq:delay_2}
V_{b}(\nu )=\int_{sky}A(\widehat{s},\nu )I(\widehat{s},\nu )e^{-i2\pi\nu\frac{b\cdot \widehat{s}}{c}}
\end{equation}

\begin{equation}\label{eq:delay_3}
P(\textup{k})\approx \frac{X^{2}Y}{4k_{B}^{2}}\left [ \frac{{}\widetilde{V}_{b}^{2}(\tau)}{\Omega_{b}B/\lambda ^{4}} \right ] 
\end{equation}
The delay transform has therefore been used as a metric to characterize the response of the HERA dish \citep{Ewall-Wice2016,Neben2016,Thyagarajan2016,Fagnoni2016}. 

Given its very compact configuration, HERA achieves its highest sensitivity on a 14.6~m baseline and, in order to suppress foregrounds to below the EoR level at the corresponding $k$-mode, requires high suppression at high delays \citep{DeBoer2016}. \citet{Ewall-Wice2016} and \citet{Thyagarajan2016} show that the current dish+feed response suppresses foregrounds below the EoR signal at $\sim 300$~ns; access to smaller $k$ modes, however, can still be achieved using an optimal weighting scheme \citep{LiuTegmark2012,Ali2015}.
Although the SKA will adopt a different array configuration, motivated by imaging requirements that need a more random $uv$~coverage, the delay spectrum metric can be used to characterize its antenna response too. 

\begin{figure*}
\centering
\includegraphics[width=7in]{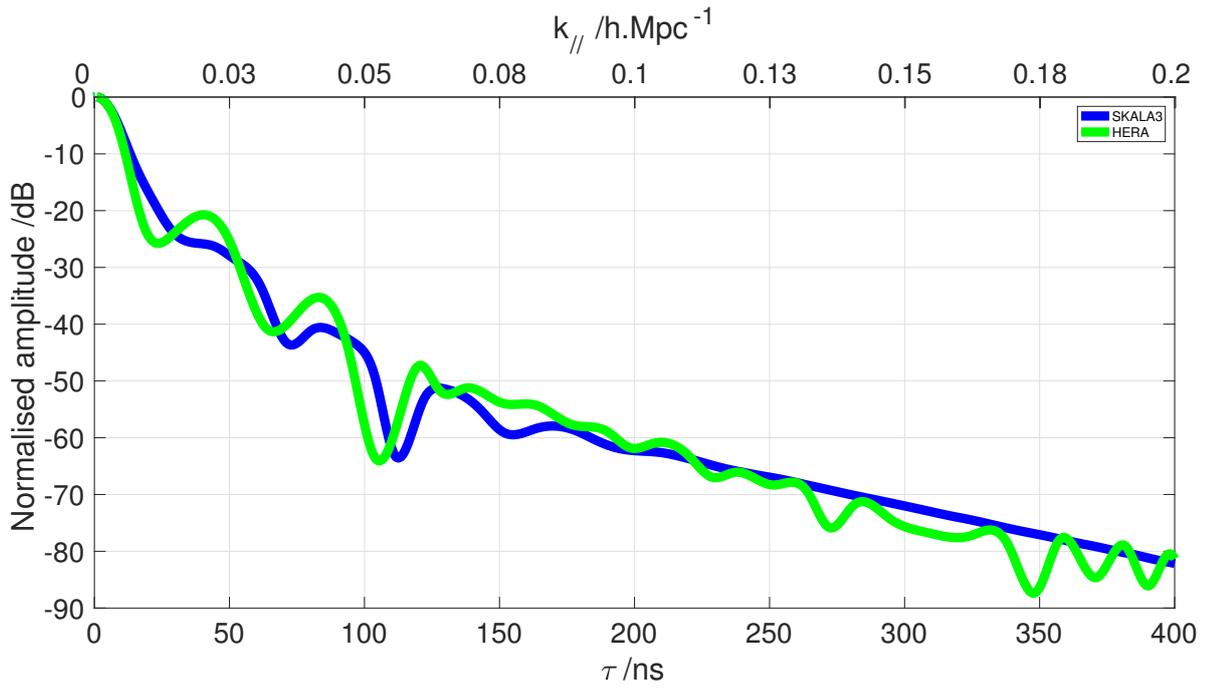}
\caption{Comparison between the normalized voltage transfer function of the HERA antenna and SKALA-3 at zenith (including in both cases the effects of their matching to the LNA) in the band 100-200 MHz after applying a Blackman-Harris window (100-200 MHz) to both of them. This window is used to eliminate the noise caused by numerical artifacts in the calculation of the transfer function towards the edges of the band.} 
\label{fig:HvS200}
\end{figure*}

\begin{figure*}
\centering
\includegraphics[width=7in]{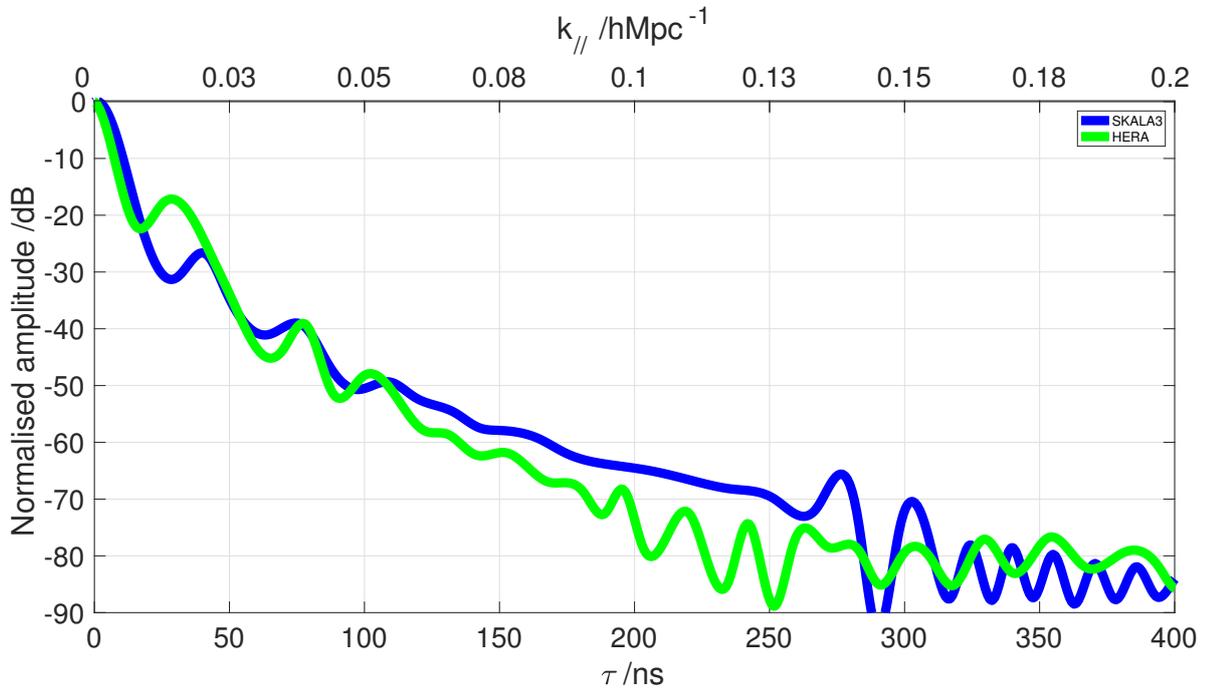}
\caption{Comparison between the normalized voltage transfer function of the HERA antenna and SKALA-3 at zenith (including in both cases the effects of their matching to the LNA) in the band 50-250 MHz after applying a Blackman-Harris window (50-250 MHz) to both of them. This window is used to eliminate the noise caused by numerical artifacts in the calculation of the transfer function towards the edges of the band.} 
\label{fig:HvS250}
\end{figure*}

\highlighttext{\citet{Ewall-Wice2016} describes} how to perform a simulated analysis of the antenna system to measure its capabilities for EoR detection using the Delay Spectrum technique. We have followed a similar approach here but \highlighttext{extended} to include the effects of the matching to the LNA. The design goal is to reduce the response of the antenna in the delay spectrum for large values of delay (corresponding to large values of $k_\|$) to avoid contamination of the EoR window. 
The core EoR detection will be done \highlighttext{using} $k_\|$ modes greater than 0.1 h/Mpc and therefore we should make sure that any power introduced by reflections in our instrument is below the EoR HI signal. 
A plane wave incoming from zenith is used to excite the HERA antenna. The same process is used to excite the SKALA-3 antenna. Circuit models of the LNAs for both HERA and SKA have been connected to the antennas in the simulation to represent the effects of the mismatch between the antenna and the LNA in their response. Separate time domain full electromagnetic simulations (using the CST program) were then performed to obtain the output signal of the pair antenna+LNA for both systems. The transfer function of the antenna is then calculated by dividing in Fourier space the output voltage signal by the input signal (the plane wave) used as excitation. Finally the calculated transfer function is transformed to the delay domain by applying an inverse Fourier transform. Figures ~\ref{fig:HvS200} and ~\ref{fig:HvS250} show this transfer function for both antennas in two frequency bands relevant to the CD/EoR science, 100-200 MHz and the extended band 50-250 MHz. These plots are normalized to the amplitude of the first incidence of the plane wave in the HERA feed after an initial reflection in the dish and to the direct incidence in the SKALA-3 antenna respectively. These incidence times are used as $\tau$ = 0 ns. 

In Figure~\ref{fig:HvS200} we can see that the SKALA-3 performance is comparable to that of the HERA antenna including feed and dish \highlighttext{for} a large portion of significant delays for the band 100-200 MHz. In Figure~\ref{fig:HvS250} we can see how the response of SKALA-3 \highlighttext{also seems to be competitive in} the extended band (50-250 MHz). In these simulations the effects of the multiple reflections on the HERA dish are included. This result is likely to change when the effects of mutual coupling are included for both systems. \hltr{In the case of SKA1-LOW the antennas in a station will be arranged in a pseudo-random configuration and they will be placed at about 2 meters from each other on average. However for HERA the configuration is triangular-regular and the distance between antennas is 14.6 m. While the level of the coupling will be therefore higher for SKA, it is also worth noticing that the effects of mutual coupling tend to average out in random arrays \citep{deLera2011b}. The actual impact of coupling could therefore be assumed to be non dominant, but further work will explore these effects in more detail.} The HERA results shown here are \highlighttext{consistent} with those shown in \citet{Ewall-Wice2016} although the results in this paper include the effects the mismatch with the actual LNA. 

It is also worth noting that HERA \highlighttext{only uses} relative\hltr{ly} short baselines to prevent the contamination of the EoR window.

\section{Conclusions and future work}
\label{sec:conclusions}

In this paper we have presented two different analyses of the passband response of \highlighttext{realistic antennas for 21-cm cosmology experiments.} One is based on fitting a local low-order polynomial to the passband in order to model and calibrate it. The second is based on \highlighttext{building} an antenna design free of spectral structure at the relevant scales for the detection of the EoR following the Delay Spectrum technique proposed for HERA. \highlighttext{The analyses have been done using the improved log-periodic antenna proposed for SKA1-LOW (SKALA-3).} 

\highlighttext{Using electromagnetic and RF computer simulations of the antennas and LNAs,} we have shown that with minor modifications \highlighttext{on the antenna and LNA design it is possible to improve the impedance matching that dominates the spectral performance of the system meeting the passband smoothness requirements} established in \citet{Trott2016}. Furthermore, \highlighttext{these modifications result in a design} able to use the Delay Spectrum technique provided that other considerations are also taken into account (e.g. the effects of mutual coupling and side-lobes have not been taken into account in this analysis). The main parameters of the antenna such as sensitivity \highlighttext{are} only slightly affected by these changes\highlighttext{, however, the overall response has been flattened at reduced spectral scales over the frequency range.}

\highlighttext{These analyses are} now going to be validated using measurements \highlighttext{on} a prototype. Furthermore, we will shortly add to \highlighttext{the} study an analysis of the effects of mutual coupling and other sources of spectral features in the instrument's passband (e.g. side-lobes as in \cite{Thyagarajan2016}). \hltr{In particular, beam-formed arrays of antennas are studied in paper II in this series \citep{trottskala17}. In \citet{trottskala17} realistic interferometric and sky simulations (including frequency-dependent primary beam shapes and array configuration) are used} to study the calibration performance of the antennas. It applies a weighted least squares polynomial estimator to assess the precision with which each antenna type can calibrate the instrument, and compares \highlighttext{to} the tolerances described in \citet{Trott2016}. \hltr{Further work will also be done to assess the impact of the spectral smoothness of realistic beam-forming technologies and techniques.} 

\section*{Acknowledgements}
The authors thank Paul Alexander and the SKA Office for useful discussions. The authors would like to acknowledge their SKA Aperture Array Design Consortium colleagues for their help and discussions on the topic. This research was supported by the Science \& Technology Facilities Council (UK) grant: \textit{SKA, ST/M001393/1} and the University of Cambridge, UK. This research was also supported under the Australian Research Council's Discovery Early Career Researcher funding scheme (project number DE140100316), and the  Centre for All-sky Astrophysics (an Australian Research Council Centre of Excellence funded by grant CE110001020). This work was supported by resources provided by the Pawsey Supercomputing Centre with funding from the Australian Government and the Government of Western Australia. We acknowledge the iVEC Petabyte Data Store, the Initiative in Innovative Computing and the CUDA Center for Excellence sponsored by NVIDIA at Harvard University, and the International Centre for Radio Astronomy Research (ICRAR), a Joint Venture of Curtin University and The University of Western Australia, funded by the Western Australian State government. \hltr{The authors also acknowledge support from the Royal Society and the Newton Fund under grant NA150184. This work is based on the research supported in part by the National Research Foundation of South Africa (Grant Number 103424).}




\bibliographystyle{mnras}
\bibliography{SKALApaperI} 







\bsp	
\label{lastpage}
\end{document}